\documentclass[twocolumn,prc,amsmath,amssymp,showpacs,showkeys,nofootinbib]{revtex4}
\usepackage{bm} 

\newcommand{\ri}{\mathrm{i}}
\newcommand{\re}{\mathrm{e}}
\newcommand{\rd}{\mathrm{d}}
\newcommand{\rs}{\text{s}}
\newcommand{\bk}{{\bm{k}}}
\newcommand{\br}{{\bm{r}}}
\newcommand{\bq}{{\bm{q}}}
\newcommand{\bz}{{\bm{0}}}
\newcommand{\bn}{{\bm{n}}}
\renewcommand{\ao}{a^{\phantom\dag}}
\newcommand{\aoa}{a^{\dag}}
\newcommand{\bo}{b^{\phantom\dag}}
\newcommand{\boa}{b^{\dag}}
\newcommand{\tbo}{\widetilde{b}^{\phantom\dag}}
\newcommand{\tboa}{\widetilde{b}^{\dag}}
\newcommand{\bv}{{\bm{v}}}
\newcommand{\bb}{{\bm{b}}}
\newcommand{\ba}{{\bm{a}}}

\newcommand{\tr}{{\text{t}}}
\newcommand{\sign}{{\rm sign}}

\begin{document}


\title{Ground-state energy and depletions for a dilute binary Bose gas}

\author{Andr\'{e} Eckardt} 
\email{eckardt@theorie.physik.uni-oldenburg.de}
\author{Christoph Weiss}
\author{Martin Holthaus}

\affiliation{Institut f\"ur Physik, Carl von Ossietzky Universit\"at, 													D-26111 Oldenburg, Germany} 
\date{\today} 

\begin{abstract}
When calculating the ground-state energy of a weakly interacting Bose gas with the 
help of the customary contact pseudopotential, one meets an artifical ultraviolet 
divergence which is caused by the incorrect treatment of the true interparticle 
interactions at small distances. We argue that this problem can be avoided by 
retaining the actual, momentum-dependent interaction matrix elements, and use this 
insight for computing both the ground-state energy and the depletions of a binary Bose 
gas mixture. Even when considering the experimentally relevant case of equal masses 
of both species, the resulting expressions are quite involved, and no straightforward 
generalizations of the known single-species formulas. On the other hand, we demonstrate 
in detail how these latter formulas are recovered from our two-species results in the 
limit of vanishing interspecies interaction.   
\end{abstract}

 

\maketitle

\section{Introduction \label{Sec:Int}}

Recently, the experimental investigation of multicomponent Bose--Einstein 
condensates has made substantial progress~\cite{MyattEtAl97,
StamperKurnEtAl98,HallEtAl98,MatthewsEtAl99,MaddaloniEtAl00}. 
A considerable amount of theoretical work has been devoted to such systems,
focusing, among others, on mean-field descriptions of trapped binary 
mixtures~\cite{HoShenoy96,EsryEtAl97},
stability properties and phase segregation~\cite{GoldsteinMeystre97,LawEtAl97,
GoldsteinEtAl00,AoChui98,AoChui00,AlexandrovKabanov02,Yukalov04},
collective excitations~\cite{BuschEtAl97,GrahamWalls98,PuBigelow98},
spinor condensates~\cite{Ho98,LawEtAl98}, and multicomponent condensates 
in optical lattices~\cite{ChenWu03,KrutitskyGraham03}. 

In this paper we determine the ground-state energy and the depletions for a homogeneous 
binary condensate within the Bogoliubov approximation~\cite{Bogoliubov47}. The 
calculation of the ground-state energy beyond the mean-field approximation, and of the 
depletion, of a single-species condensate is a major achievement of the theory of weakly 
interacting Bose gases designed in the late fifties~\cite{LeeYang57,LeeHuangYang57,
BruecknerSawada57,Beliaev58,HugenholtzPines59,Lieb63}. Recent significant developments 
include the first mathematically rigorous proof of the leading term of the commonly 
accepted expression for the ground-state energy per particle~\cite{LiebYngvason98},
the explicit calculation of nonuniversal contributions due to three-body 
scattering within the framework of effective field 
theory~\cite{BraatenNieto99,BraatenEtAl01,Andersen04}, and the clarification of 
the relation between the Gross--Pitaevskii and Bogoliubov descriptions of a
dilute Bose gas~\cite{Leggett03}. One of the technical difficulties presented by the 
subject stems from the fact that typical atomic interaction potentials exhibit a strong 
increase at short distances, so that their Fourier transforms diverge for vanishing 
momentum transfer. Hence, one customarily replaces the true interaction potential by 
the contact pseudopotential 
\begin{equation}\label{Eq:PseudoS}
    U_\text{p}(\br) = \frac{4\pi a\hbar^2}{m} \delta(\br) \; ,
\end{equation}  
where $m$ denotes the atomic mass, and $a$ is the $s$-wave scattering length
for collisions among the cold atoms. Thus, one relies on the knowledge that 
the low-temperature properties of dilute Bose gases are governed solely 
by the scattering lengths, an assertion commonly referred to as the 
``Landau postulate''~\cite{ChernyShanenko00}. Indeed, the leading terms
of an expansion of the energy density for the ground state of a dilute 
homogeneous single-species gas of Bose particles interacting via some 
potential with scattering length~$a$ can be written as~\cite{Leggett03}
\begin{equation}\label{Eq:LY}
   \frac{E_0}{V} = \frac{2\pi a \hbar^2}{m} n^2 
   \left[ 1 + \frac{128}{15\sqrt{\pi}} \sqrt{n a^3} \right] \; ,  
\end{equation}
first derived by Lee and Yang for the particular case of hard-sphere
Bosons with diameter~$a$~\cite{LeeYang57,LeeHuangYang57}. Nonuniversal,
{\em i.e.\/} potential-dependent contributions figure only as higher-order
corrections, due to the effective range for $s$-wave scattering, and 
three-body interaction~\cite{BraatenNieto99,BraatenEtAl01}.  

While the standard contact pseudopotential~(\ref{Eq:PseudoS}) is sufficient 
for determining the quasiparticle energy dispersion for a weakly interacting Bose 
gas, problems arise in the calculation of its ground-state energy. Namely, when 
coupling bare-particles states with opposite momenta $\hbar\bk$ and $-\hbar\bk$ 
for setting up the harmonic oscillator-like states corresponding to the phonon 
quasiparticles, one also has to account for the ``vacuum'' energies of these 
oscillators. The energy of the ground state of the weakly interacting gas, 
defined by the absence of phonons for all~$\bk$, then acquires a contribution 
from the sum over all these vacuum energies. However, if one naively uses 
the pseudopotential~(\ref{Eq:PseudoS}), all interaction matrix elements are 
replaced by a constant, instead of being properly suppressed at high momentum
transfer, so that this sum becomes ultra\-violet divergent. It is understood that 
this divergence is not a fundamental physical problem, but rather a consequence 
of the incorrect handling of the true interparticle interaction at short
distances through the delta-function pseudopotential~(\ref{Eq:PseudoS}): Actually, 
the proper pseudo\-potential operator derived by Huang and Yang~\cite{HuangYang57}
is given by
\begin{equation}\label{Eq:PseudoR}
    U_\text{p}(\br) = \frac{4\pi a\hbar^2}{m} \delta(\br)
    \frac{\partial}{\partial r}r \; ;
\end{equation}  
when operating on wave functions with a $1/r$-type singularity, this operator yields 
results quite different from those given by its simplification~(\ref{Eq:PseudoS}). 
The modern solution to the problem is based on local effective field theory and
renormalization theory, as reviewed recently in Ref.~\cite{Andersen04}.
  
A possible strategy for avoiding such complications has been suggested by Lifshitz 
and Pitaevskii~\cite{Landau-Stat2}: If one accepts the Landau postulate, one may 
replace the true, singular interparticle interaction potential by a smooth, 
well-behaved one, with the requirement that the substitute shares the same value of 
the scattering length~$a$. If then the result of the calculations, performed with 
the auxiliary potential, can be expressed entirely in terms of~$a$, this result is 
the same as the one provided by the original interaction. 

In the following sections, we exploit this idea for computing the ground-state 
energy and the depletions of a binary condensate mixture by considering soft, 
nonsingular interaction potentials. Contrary to standard textbook 
approaches~\cite{Landau-Stat2,Pathria-Stat}, we do not replace the interaction
matrix elements by constants, but keep track of their actual momentum dependence,
so that the spurious ultraviolet divergence is avoided altogether. Only at the end 
of the calculation, we identify the various contributions to the ground-state energy 
of the binary gas with certain terms of the Born series for the three scattering 
lengths involved. This allows us to cast our results into the desired form, enabeling 
a comparison with the known single-species results. Thus, we do neither invoke the 
pseudopotential~(\ref{Eq:PseudoS}) nor its refined form~(\ref{Eq:PseudoR}), but
work directly with nonsingular interaction potentials. 

We proceed as follows: In Sec.~\ref{Sec:Hamiltonian} the quadratic 
low-temperature approximation to the full Hamiltonian of the binary gas is 
obtained by adapting the reasoning customarily applied in the single-species 
case. In Sec.~\ref{Sec:Diagonalisation} this approximate Hamiltonian 
is then diagonalized exactly, relying on a method used by Hopfied in 1958 
when studying the interaction of the radiation field with a polarization
field~\cite{Hopfield58}. In this way, we necessarily recover the already
known results for the quasiparticle excitation 
spectra~\cite{GoldsteinMeystre97,AlexandrovKabanov02}, but we also
identify an energy shift associated, for each wavevector~$\bk$, with 
the transformation to the quasiparticle form of the Hamiltonian. In order
to obtain the ground-state energy of the system, the sum over all these
shifts has to be evaluated; this is achieved in Sec.~\ref{Sec:GrStEn} 
with the help of a density expansion which yields only finite quantities at 
each intermediate step of the calculation~\cite{WeissEtAl04}. The depletions for 
a binary condensate are determined by the coefficients of the quasiparticle 
transformation, and calculated at the end of Sec.~\ref{Sec:GrStEn}. In 
Sec.~\ref{Sec:Conclusion} we briefly discuss our main findings.

\section{The Hamiltonian in Quadratic Approximation \label{Sec:Hamiltonian}}

The Hamiltonian of a homogeneous gas consisting of a mixture of $N_i$ Bose particles 
of species labeled by $i$ (with $i = 1,2,\ldots$) in a cubic volume~$V$ reads  
\begin{eqnarray}\label{Eq:H-Psi}
    H &=& \sum_i \, \int_V \! \rd \br \,
    \Psi_i^\dag(\br) h_i(\br) \Psi_i^{}(\br)  
    \nonumber\\&&
    +\frac{1}{2} \sum_{i,j} \, \int_V \int_V \! \rd \br_1 \, \rd \br_2 \,
    \Psi_i^\dag(\br_1) \Psi_j^\dag(\br_2)
    \nonumber\\&&\qquad\qquad\qquad\times
    U_{ij}(\br_1-\br_2) \, \Psi_j(\br_2)\Psi_i(\br_1) \; ,
\end{eqnarray}
where $U_{ij}(\br) = U_{ji}(\br)$ denotes the real, isotropic interaction 
potential between particles of the type~$i$ and those of type~$j$. We assume
that the intraspecies potentials $U_{ii}(\br)$ be repulsive; none of the potentials 
should admit any many-body bound state. The single-particle Hamiltonian $h_i(\br)$ 
is given by the kinetic energy operator for free particles of mass~$m_i$, 
$h_i(\br) = -(\hbar^2/2m_i)\Delta_\br$. The thermodynamic limit will be taken at 
a later stage, $N_i \to \infty$ and $V \to \infty$ such that the density 
$n_i = N_i/V$ of each species remains constant. The $\Psi_i^{}(\br)$ are the usual 
Bose field operators, obeying the commutation relations
\begin{eqnarray}\label{Eq:ComRelPsi}
    &&\big[\Psi_i^{}(\br),\Psi_j^\dag(\br')\big] = \delta_{i,j} \, \delta(\br-\br') \; ,
    \nonumber\\
    &&\big[\Psi_i^{}(\br),\Psi_j^{}(\br') \big]
    = \big[\Psi_i^\dag(\br),\Psi_j^\dag(\br')\big] = 0 \; .
\end{eqnarray}
Imposing periodic boundary conditions, the single-particle eigenfunctions of $h_i(\br)$ 
are given by plane waves, $\varphi_\bk(\br)= V^{-1/2}\exp(\ri\bk\br)$, labeled by 
wavevectors $\bk=2\pi V^{-1/3}\bn$, where $\bn$ is a vector with integer components. 
The field operators then are expanded in this basis, 
\begin{equation}\label{Eq:Psi-a}
    \Psi_i^{}(\br) = \sum_\bk a^{}_{i\,\bk} \varphi^{}_\bk(\br) \; .
\end{equation}   
The coefficients $a_{i\,\bk}$ and their adjoints $\aoa_{i\,\bk}$ are the familiar Bose 
annihilation and creation operators for particles of species~$i$ in the single-particle 
state~$\bk$. As a consequence of Eq.~(\ref{Eq:ComRelPsi}), they satisfy the commutation
relations 
\begin{eqnarray}\label{BinComRela}
    &&\big[\ao_{i\,\bk}\, , \aoa_{j\,\bk'}\big]
    = \delta_{i,j} \, \delta_{\bk,\bk'} \; ,
    \nonumber\\
    &&\big[\ao_{i\,\bk}\, , \ao_{j\,\bk'}\big]
    =\big[\aoa_{i\,\bk}\, , \aoa_{j\,\bk'}\big]
    = 0 \; .
\end{eqnarray}
Introducing the Fourier transforms of the interaction potentials,
\begin{eqnarray}\label{Eq:FourierComp}
    u_{ij}(\bq) \equiv \int_V \! \rd\br \, \re^{-\ri\bq\br} U_{ij}(\br) \; ,
\end{eqnarray} 
and inserting the expansion~(\ref{Eq:Psi-a}) into the Hamiltonian~(\ref{Eq:H-Psi}),
one obtains
\begin{eqnarray}\label{Eq:OrigH}
    H &=&
    \sum_i \, \sum_{\bk} \frac{\hbar^2\bk^2}{2m_i} \,
    \aoa_{i\,\bk} \ao_{i\,\bk}
    \nonumber\\&&
	+\sum_{i,j}{\sum_{\{\bk\}}}'\;
    \frac{u_{ij}(\bk_1-\bk_1')}{2V} \,
    \aoa_{i\,\bk_1}\aoa_{j\,\bk_2}\ao_{j\,\bk'_2}\ao_{i\,\bk'_1} \;, 
\end{eqnarray}
where the prime at the summation sign indicates restriction to those terms which
comply with momentum conservation, $\bk_1+\bk_2 = \bk'_1+\bk'_2$. Note that
$u_{ij}(\bq)$ carries the dimension of energy times volume; moreover, we have 
$u_{ij}(-\bq)=u_{ij}(\bq)$.

For interaction potentials with a strong short-range singularity, such as 
Lennard-Jones or hard-sphere potentials, the Fourier components (\ref{Eq:FourierComp}) 
diverge for $\bq \to \bz$; this divergence can be seen as a consequence of the use 
of a single-particle basis $\varphi_\bk(\br)$. Therefore, keeping in mind the 
remarks made in the introduction, we perform the following analysis with ``soft''
interactions for which the Fourier transforms remain finite, while stipulating 
that their scattering lengths be the same as those of typical atomic 
potentials~\cite{Landau-Stat2}. 

Under the assumption that there exists a condensate for each particle species,
such that the single-particle ground state is occupied by macroscopically large 
numbers of particles of both types, we employ the conventional quadratic 
approximation for the description of the low-energy 
dynamics~\cite{Bogoliubov47,Landau-Stat2,Pathria-Stat}: Firstly, the interaction 
terms of the Hamiltonian are categorized with respect to the number of ground-state 
operators $a_{i\,\bz}$ or $a^{\dag}_{i\,\bz}$ they contain. The relevant matrix 
elements of these operators should be roughly on the order of $N_i^{1/2}$, and thus 
large compared to the corresponding matrix elements of the operators $a_{i\,\bk}$ or 
$a^{\dag}_{i\,\bk}$ referring to excited states, and to sums over such elements. 
However, it will become obvious in the following that this handwaving reasoning 
requires some  caution. Secondly, the ``off-diagonal'' operators $a_{i\,\bz}$ and 
$a^{\dag}_{i\,\bz}$ are replaced by the square roots $N_{i\,\bz}^{1/2}$ of the
single-particle ground-state occupation numbers. This is acceptable as long as the 
probability distributions for finding $N_{i\,\bz}$ particles in the ground state 
are broad compared to~$1$, so that shifts by one particle do not essentially change 
the many-particle ground state. Neglecting supposedly small contributions to the 
Hamiltonian with three or more nonzero wavenumber indices, one finds
\begin{eqnarray} 
    &&{\sum_{\{\bk\}}}'\;u_{ij}(\bk_1-\bk_1') \,
    \aoa_{i\,\bk_1}\aoa_{j\,\bk_2}\ao_{j\,\bk'_2}\ao_{i\,\bk'_1}
    \nonumber \\ & \simeq & 
    u_{ij}(\bz) \, N_{i\,\bz}N_{j\,\bz} 
    \nonumber\\ & + & \sum_{\bk\ne\bz} \Big[ 
      u_{ij}(\bk) \, \aoa_{i\,\bk}\aoa_{j\,-\bk} \, \sqrt{N_{i\,\bz}N_{j\,\bz}} 
    \nonumber \\ && \quad \;\;
    + u_{ij}(\bk) \, \ao_{i\,\bk} \ao_{j\,-\bk}  \, \sqrt{N_{i\,\bz}N_{j\,\bz}}
    \nonumber \\ && \quad \;\;	
    + u_{ij}(\bk) \, \aoa_{i\,\bk}\ao_{j\,\bk}   \, \sqrt{N_{i\,\bz}N_{j\,\bz}}
    \nonumber \\ && \quad \;\;
    + u_{ij}(\bk) \, \aoa_{j\,\bk}\ao_{i\,\bk}   \, \sqrt{N_{i\,\bz}N_{j\,\bz}}
    \nonumber \\ && \quad \;\;
    + u_{ij}(\bz) \, \aoa_{i\,\bk}\ao_{i\,\bk}   \, N_{j\,\bz}
    \nonumber \\ && \quad \;\;	
    + u_{ij}(\bz) \, \aoa_{j\,\bk}\ao_{j\,\bk}   \, N_{i\,\bz} \Big] \; . 	
\end{eqnarray}
Now the relations 
$N_{i\,\bz} =  N_i - \sum_{\bk\ne\bz}\aoa_{i\,\bk}\ao_{i\,\bk}$ can be employed 
to express the ground-state occupation numbers~$N_{i\,\bz}$ in terms of the 
total particle numbers~$N_i$ of each species. Again neglecting terms which
are considered as small within the above categorization scheme, this leads to the 
cancelation of contributions with zero momentum transfer:      
\begin{eqnarray}
    &&{\sum_{\{\bk\}}}'\;u_{ij}(\bk_1-\bk_1') \,
    \aoa_{i\,\bk_1}\aoa_{j\,\bk_2}\ao_{j\,\bk'_2}\ao_{i\,\bk'_1} 
    \nonumber\\ & \simeq & 
    u_{ij}(\bz) N_iN_j + \sum_{\bk\ne\bz} u_{ij}(\bk) \sqrt{N_iN_j} 
    \nonumber \\ && \times
    \Big[ \aoa_{i\,\bk}\aoa_{j\,-\bk}
         +\ao_{i\,\bk} \ao_{j\,-\bk}
         +\aoa_{i\,\bk}\ao_{j\,\bk}
         +\aoa_{j\,\bk}\ao_{i\,\bk} \Big] \; .
\end{eqnarray}
Restricting ourselves from here on to only two particle species $1$ and $2$,
within the quadratic approximation the low-energy Hamiltonian takes the form
\begin{equation}\label{Eq:HQA}
    H_{\text{QA}} = A_\bz+{\sum_{\bk\ne\bz}}^{2\pi}  H_\bk \; ,
\end{equation}
where the superscript $2\pi$ at the summation sign is meant to indicate 
restriction of the sum to one $\bk$-halfspace, and 
\begin{eqnarray}\label{Eq:H_ABCD}
    H_\bk & = & \sum_{i=1}^2  \left[
    B_{i\,\bk} \left(\aoa_{i\,\bk}\aoa_{i\,-\bk} +\ao_{i\,\bk}\ao_{i\,-\bk} \right)
    + C_{i\,\bk} \, \aoa_{i\,\bk}\ao_{i\,\bk}   \right]
    \nonumber\\
    & + & D_\bk \left(\aoa_{1\,\bk}\aoa_{2\,-\bk}+\ao_{1\,\bk}\ao_{2\,-\bk}
			       +\aoa_{1\,\bk}\ao_{2\,\bk}+\ao_{1\,\bk}\aoa_{2\,\bk}\right)
    \nonumber\\
    & + & \left(\; \bk \, \rightarrow -\bk \;\right) \; .
\end{eqnarray}
The symbol $\left(\; \bk \, \rightarrow -\bk \;\right)$ denotes contributions of the
same kind as appearing in the first two lines of this definition, with $\bk$ replaced 
by $-\bk$. Moreover, we have introduced the abbreviations
\begin{subequations}\label{Eq:ABCD}
\begin{eqnarray}
    A_\bz &=& \frac{1}{2}v_{11}(\bz)N_1 +\frac{1}{2} v_{22}(\bz)N_2 
    \label{Eq:ABCD1}
    \nonumber \\ &&
              +v_{12}(\bz)\sqrt{N_1N_2}	\; ,		  
    \\
    B_{i\,\bk}&=&B_{i\,-\bk} = \frac{1}{2} v_{ii}(\bk)  \;,
    \\
    C_{i\,\bk}&=&C_{i\,-\bk} = t_i(\bk) + v_{ii}(\bk)  \;,
    \\
    D_\bk&=&D_{-\bk} \;= v_{12}(\bk)  \; ,
\end{eqnarray}
\end{subequations}
with
\begin{subequations}\label{Eq:tivij}
\begin{eqnarray}
	t_i(\bk)	&\equiv&	\frac{\hbar^2\bk^2}{2m_i} \; ,
	\\
	v_{ij}(\bk)	&\equiv&	\sqrt{n_in_j}u_{ij}(\bk)  \; ,  	
\end{eqnarray}
\end{subequations}
which will be used extensively in the following. Thus, within the quadratic 
approximation the Hamiltonian of the Bose gas mixture is given by a sum of 
commuting parts $H_\bk$, quadratic in the $a$-operators, which can be diagonalized 
separately; the interactions couple particles of both types with momenta 
$\hbar\bk$ and $-\hbar\bk$.

\section{Quasiparticle form of the Hamiltonian \label{Sec:Diagonalisation}}

In order to convert $H_\bk$ into a form which corresponds to noninteracting
quasiparticles, we consider excitations characterized by a wavevector~$\bk \ne \bz$.
Such excitations are described by linear combinations of annihilation operators 
for particles with momentum $\hbar\bk$ and creation operators for particles with 
reverse momentum $-\hbar\bk$,   
\begin{equation}\label{Eq:Ansatz}
    \tbo_{i\,\bk} = u_{i\,\bk}\ao_{1\,\bk}   + v_{i\,\bk} \aoa_{1\,-\bk}
               	  + \mu_{i\,\bk}\ao_{2\,\bk} + \nu_{i\,\bk} \aoa_{2\,-\bk} \; ,
\end{equation}
so that $u_{i\,\bk}$ and $v_{i\,\bk}$ quantify the contributions of species~$1$
to the $i$th quasiparticle excitation, and $\mu_{i\,\bk}$ and $\nu_{i\,\bk}$
those of species~$2$. As a consequence of the symmetry of the 
coefficients~(\ref{Eq:ABCD}), we also have $u_{i\,\bk} = u_{i\,-\bk}$, {\em etc.\/} 
We then require the commutation relations~\cite{Hopfield58}
\begin{eqnarray}\label{Eq:Demand}
    \Big[\tbo_{i\,\bk}\, , H_\bk\Big] = \widetilde{\varepsilon}_i(\bk) \tbo_{i\,\bk} \; ,
\end{eqnarray}
in analogy to the relation fulfilled by the Hamiltonian and the annihilation operator
of the simple harmonic oscillator. Inserting the ansatz~(\ref{Eq:Ansatz}) and the 
expression~(\ref{Eq:H_ABCD}) for $H_\bk$ into these relations, utilizing the symmetry 
of the coefficients under interchange of $\bk$ and $-\bk$, and comparing cofficients 
of identical operators, leads us to the eigenvalue problem  
\begin{equation} \label{Eq:EigenProblem}
    Q_\bk \bv_{i\,\bk} = \widetilde{\varepsilon}_i(\bk) \bv_{i\,\bk} \; ,  
\end{equation}
where the $4 \times 4$ matrix $Q_\bk$ is given by
\begin{equation}\label{Eq:Qk}
    Q_\bk \equiv \left(\begin{array}{cccc}
            C_{1\,\bk} & -2B_{1\,\bk} & D_\bk & -D_\bk \\
            2B_{1\,\bk} & -C_{1\,\bk} & D_\bk & -D_\bk \\
            D_\bk & -D_\bk & C_{2\,\bk} & -2B_{2\,\bk} \\
            D_\bk & -D_\bk & 2B_{2\,\bk} & -C_{2\,\bk}
             \end{array}\right) \; ,
\end{equation}    
and 
\begin{equation}
	\bv_{i\,\bk} \equiv \left(\begin{array}{cccc}
	u_{i\,\bk} & v_{i\,\bk} & \mu_{i\,\bk} & \nu_{i\,\bk}
	                          \end{array}\right)^\tr \;.
\end{equation}
Since $Q_\bk$ is not hermitian, the eigenvalues $\widetilde{\varepsilon}_i(\bk)$ are 
not necessarily real. For continuing the analysis, we first assume that only real 
eigenvalues occur, so that also the components of the eigenvectors $\bv_{i\,\bk}$ 
can be chosen as real numbers, and will state the physical condition for the reality
of the eigenvalues later (cf.\ Eqs.~(\ref{Eq:SingSpecStabRel}) 
and~(\ref{Eq:SpecStabRel})).

The specific structure of $Q_\bk$ can be exploited with the help of the
matrix
\begin{equation}
    h \equiv \left(\begin{array}{cccc}
         0&1&0&0 \\ 1&0&0&0 \\ 0&0&0&1 \\ 0&0&1&0
                   \end{array}\right) \; :
\end{equation}
Since $h^2 = 1$, and $h Q_\bk h = -Q_\bk$, multiplying the eigenvalue
problem~(\ref{Eq:EigenProblem}) from the left by~$h$ leads to
\begin{equation}
   Q_\bk h \bv_{i\,\bk} = -\widetilde{\varepsilon}_i(\bk) h \bv_{i\,\bk} \; .
\end{equation}
Hence, if 
$\bv_{\bk} = \left(u_{\bk} \, v_{\bk} \, \mu_{\bk} \, \nu_{\bk}\right)^\tr$ 
is an eigenvector with eigenvalue $+\widetilde{\varepsilon}(\bk)$, then
$h\bv_{\bk} = \left(v_{\bk} \, u_{\bk} \, \nu_{\bk} \, \mu_{\bk}\right)^\tr$ 
is an eigenvector with eigenvalue $-\widetilde{\varepsilon}(\bk)$. Therefore, 
the eigenvalues occur in pairs of opposite sign: We may fix the indices such
that 
\begin{eqnarray}
   \widetilde{\varepsilon}_3(\bk) & = & -\widetilde{\varepsilon}_1(\bk)
   \nonumber \\ 
   \widetilde{\varepsilon}_4(\bk) & = & -\widetilde{\varepsilon}_2(\bk) \; ,
\end{eqnarray}
while stipulating that $\widetilde{\varepsilon}_1(\bk)$ and 
$\widetilde{\varepsilon}_2(\bk)$ be positive.

Introducing the further matrix 
\begin{equation}
    g \equiv \left(\begin{array}{cccc}
         1&0&0&0 \\ 0&-1&0&0 \\ 0&0&1&0 \\ 0&0&0&-1
                   \end{array}\right) \; ,
\end{equation}
which obeys $g=g^\tr=g^{-1}$, it is easy to see that $gQ_\bk$ {\em is\/} hermitian.
Hence, two eigenvectors belonging to different eigenvalues of the 
problem~(\ref{Eq:EigenProblem}) are orthogonal with respect to the bilinear form 
mediated by $g$:
\begin{eqnarray}
    \Big[\tbo_{i\,\bk}\,,\tboa_{j\,\bk}\Big] & = &
    \bv_{i\,\bk}^\tr g \bv_{j\,\bk} 
    \nonumber\\ & = &  
    0 \qquad \mbox{for} \; i \ne j \; ,
\end{eqnarray}
where the first equality is an immediate consequence of the definitions~(\ref{Eq:Ansatz})
and the commutation relations~(\ref{BinComRela}) obeyed by the bare particle operators.  
In addition, some lines of elementary algebra reveal that the eigenvalues 
$\widetilde{\varepsilon}_i(\bk)$ of the matrix $Q_\bk$ are real --- which we do 
assume here --- if and only if the matrix $g Q_\bk$ is positive semi\-definite. 
Hence, we also have 
\begin{equation}
   \widetilde{\varepsilon}_i(\bk) \bv_{i\,\bk}^\tr g \bv_{i\,\bk}
   = \bv_{i\,\bk}^\tr gQ_\bk\bv_{i\,\bk} \ge0 \; . 
\end{equation}
Therefore, normalizing the eigenvectors $\bv_{i\,\bk}$ such that 
$|\bv_{i\,\bk}^\tr g\bv_{i\,\bk}| = 1$ (note that it is not possible to change the sign 
of $\bv_{i\,\bk}^\tr g\bv_{i\,\bk}$ by normalization), we thus find that the sign of 
$\bv_{i\,\bk}^\tr g\bv_{i\,\bk}$ is fixed by the sign of the corresponding eigenvalue:
\begin{eqnarray}\label{Eq:ONRel}
   \Big[\tbo_{i\,\bk}\,,\tboa_{j\,\bk}\Big] & = &
   \bv_{i\,\bk}^\tr g \bv_{j\,\bk} 
   \nonumber\\ & = &
   \sign(\widetilde{\varepsilon}_i(\bk)) \, \delta_{ij} \; .
\end{eqnarray} 

Multiplying the basic commutation relation~(\ref{Eq:Demand}) from the left by 
$\tboa_{i\,\bk}$, then multiplying its adjoint relation from the right by
$\tbo_{i\,\bk}$, and subtracting, immediately gives 
\begin{equation}
    \Big[\tboa_{i\,\bk}\tbo_{i\,\bk}\, , H_\bk\Big] = 0 \; ,
\end{equation}
again relying on the proposition that $\widetilde{\varepsilon}_i(\bk)$ be real. 
Since $H_\bk$ depends quadratically on creation and annihilation operators for the 
bare particles, and the transformation~(\ref{Eq:Ansatz}) is linear, this operator 
therefore has to have the form
\begin{equation}
    H_\bk = \sum_{i=1}^4 f_{i\,\bk} \, \tboa_{i\,\bk}\tbo_{i\,\bk} + g_{\bk}  
\end{equation}
when expressed in terms of $\widetilde{b}$-operators, where $f_{i\,\bk}$ and 
$g_{\bk}$ are real numbers. Once again invoking the relation~(\ref{Eq:Demand}) 
then yields the identities 
\begin{equation}
    f_{i\,\bk} \Big[\tbo_{i\,\bk}\,,\tboa_{i\,\bk}\Big]\tbo_{i\,\bk}
    = \varepsilon_i(\bk)\tbo_{i\,\bk} \; , 
\end{equation} 
which, in the light of the commutators~(\ref{Eq:ONRel}), allows us to identify 
$f_{i\,\bk}$ with the absolute value of $\widetilde{\varepsilon}_i(\bk)$:
\begin{equation}\label{Eq:HepsG}
    H_\bk = \sum_{i=1}^4 |\widetilde{\varepsilon}_i(\bk)| \,
    \tboa_{i\,\bk}\tbo_{i\,\bk} + g_{\bk} \; .
\end{equation}

In order to exploit these algebraic deliberations for the construction of proper
quasiparticle annihilation and creation operators $\bo_{i\,\bk}$ and
$\boa_{i\,\bk}$ pertaining to the two-species Bose gas, we now have to make 
sure that these latter operators satisfy the Bose commutation relations
\begin{eqnarray}\label{Eq:bComRel}
    && \big[\bo_{i\,\bk}\, , \boa_{j\,\bk'}\big]
       = \delta_{i,j}\,\delta_{\bk,\bk'}  \;,
    \nonumber\\
    && \big[\bo_{i\,\bk}\, , \bo_{j\,\bk'}\big]
       =\big[\boa_{i\,\bk}\, , \boa_{j\,\bk'}\big]
       = 0 \; .
\end{eqnarray}
In view of the orthonormality-type relations~(\ref{Eq:ONRel}), this is easy to
achieve: The two eigenvectors $\bv_{1\,\bk}$ and $\bv_{2\,\bk}$ associated with 
positive eigenvalues $\widetilde{\varepsilon}_1(\bk)$ and 
$\widetilde{\varepsilon}_2(\bk)$ already provide a commutator with the required
magnitude and sign, and therefore directly give rise, by means of 
Eq.~(\ref{Eq:Ansatz}), to quasiparticle annihilation operators,
\begin{equation}
    \tbo_{i\,\bk} = \bo_{i\,\bk} 
    \quad \mbox{for} \; i = 1,2 \; ;
\end{equation}
we also write
\begin{equation}
    \widetilde{\varepsilon}_i(\bk) = \varepsilon_i(\bk) 
    \quad \mbox{for} \; i = 1,2 \; .
\end{equation}    
Next, inspecting  $i = j = 3$ and $i = j =4$, the other two eigenvectors 
$\bv_{3\,\bk} = h\bv_{1\,\bk}$ and $\bv_{4\,\bk} = h\bv_{2\,\bk}$ produce 
a negative sign on the r.h.s.\ of Eq.~(\ref{Eq:ONRel}). Therefore, these
eigenvectors are associated with quasiparticle creation operators; evidently,
we have
\begin{equation}
    \tbo_{3\,\bk} = \boa_{1\,-\bk} \quad , \quad 
    \tbo_{4\,\bk} = \boa_{2\,-\bk}
\end{equation}
with positive quasiparticle energies
\begin{equation}
   |\widetilde{\varepsilon}_3(\bk)| = \varepsilon_1(\bk) \quad , \quad 
   |\widetilde{\varepsilon}_4(\bk)| = \varepsilon_2(\bk) \; .
\end{equation}
Note that $\varepsilon_i(\bk) = \varepsilon_i(-\bk)$ for $i = 1,2$, as a 
consequence of the symmetry of the coefficients~(\ref{Eq:ABCD}). Finally,  
relabeling the operators in Eq.~(\ref{Eq:HepsG}) according to these
prescriptions, rearranging, and combining all emerging $c$-numbers into energy 
shifts $\alpha(\bk)$, allows us to cast the operators $H_\bk$ into the standard 
forms   
\begin{eqnarray}\label{Eq:HDem}
    H_\bk & = & \alpha(\bk) + \varepsilon_1(\bk) \boa_{1\,\bk}\bo_{1\,\bk}
        		    + \varepsilon_2(\bk) \boa_{2\,\bk}\bo_{2\,\bk}
    \nonumber \\ & & 
              + \left( \bk  \rightarrow -\bk \right) \; .
\end{eqnarray}
In this way, the above construction directly leads to the desired  oscillator 
Hamiltonian for noninteracting quasiparticles, provided we restrict ourselves to 
conditions such that the eigenvalues of the problem~(\ref{Eq:EigenProblem}) are 
real (or, equivalently, those of the matrix $gQ_\bk$ are non-negative), {\em and\/} 
we select the two positive eigenvalues $\varepsilon_1(\bk)$ and $\varepsilon_2(\bk)$: 
Although the eigenvalues of the problem~(\ref{Eq:EigenProblem}) occur with both signs, 
negative quasiparticle energies are strictly excluded by the algebra.

Of particular significance for our purposes are the constants $\alpha(\bk)$ 
occurring in the quasiparticle Hamiltonians~$H_\bk$: For each $\bk$, 
transformation of the original operator~(\ref{Eq:H_ABCD}) to the normal
form~(\ref{Eq:HDem}) necessarily involves such an energy shift. Since the 
ground state of the gas mixture is characterized by the absence of all 
quasiparticle excitations, the ground-state energy~$E_0$ is then given by the 
sum over all these shifts, plus the constant $A_\bz$ already appearing in the 
quadratic approximation~(\ref{Eq:HQA}): 
\begin{eqnarray}\label{Eq:E0Formal}
    E_0 = A_\bz + \sum_{\bk\ne\bz} \alpha(\bk) \; ,
\end{eqnarray}
with $A_\bz$ as stated in Eq.~(\ref{Eq:ABCD1}). In order to determine the
individual shifts~$\alpha(\bk)$, we substitute in Eq.~(\ref{Eq:HDem}) the 
original $a$-operators for the $b$-operators. Comparison with Eq.~(\ref{Eq:H_ABCD}) 
then yields 
\begin{equation}
    \alpha(\bk) = -\varepsilon_1(\bk)\left(v_{1\,\bk}^2+\nu_{1\,\bk}^2\right)
                  -\varepsilon_2(\bk)\left(v_{2\,\bk}^2+\nu_{2\,\bk}^2\right) \; ,
\end{equation}
together with
\begin{eqnarray}	      	  
    C_{1\,\bk} &=&  \varepsilon_1(\bk)\left(u_{1\,\bk}^2+v_{1\,\bk}^2\right)
        +\varepsilon_2(\bk)\left(u_{2\,\bk}^2+v_{2\,\bk}^2\right) \; ,
    \nonumber \\
    C_{2\,\bk} &=&  \varepsilon_1(\bk)\left(\mu_{1\,\bk}^2+\nu_{1\,\bk}^2\right)
        +\varepsilon_2(\bk)\left(\mu_{2\,\bk}^2+\nu_{2\,\bk}^2\right) \; .
\end{eqnarray}		
Finally, utilizing Eqs.~(\ref{Eq:ONRel}) in the form 
$u_{i\,\bk}^2 - v_{i\,\bk}^2 + \mu_{i\,\bk}^2 - \nu_{i\,\bk}^2 = 1$, 
we obtain the desired relation
\begin{eqnarray}\label{Eq:alpha}
    \alpha(\bk) & = & \frac{1}{2} \left(\varepsilon_1(\bk)+\varepsilon_2(\bk)
                      -C_{1\,\bk}-C_{2\,\bk} \right) 
    \nonumber\\
		& = & \frac{1}{2} \left(\varepsilon_1+\varepsilon_2 
		      -t_1-v_{11}-t_2-v_{22} \right) \; .
\end{eqnarray}
The expression on the right-hand side does no longer contain the transformation 
coefficients. Thus, for computing the ground-state energy according to
Eq.~(\ref{Eq:E0Formal}), the knowledge of these coefficients is not required.

Before computing the depletions of the single-particle ground state, we have 
to invert the previous ansatz~(\ref{Eq:Ansatz}), and to express the bare 
$a$-operators through the quasiparticle $b$-operators. Writing
\begin{eqnarray}
    \bb_\bk \equiv  \left(\begin{array}{c}
            \bo_{1\,\bk} \\ \boa_{1\,-\bk} \\
            \bo_{2\,\bk} \\ \boa_{2\,-\bk}
            \end{array}\right) \; ,
    \qquad
    \ba_\bk \equiv  \left(\begin{array}{c}
            \ao_{1\,\bk} \\ \aoa_{1\,-\bk} \\
            \ao_{2\,\bk} \\\aoa_{2\,-\bk}
            \end{array}\right) \; ,
\end{eqnarray}
the result of the above construction takes the form
\begin{eqnarray}
    \bb_\bk = V_\bk \ba_\bk \; ,
\end{eqnarray}
with the transformation matrix
\begin{eqnarray}  \label{Eq:Vk}  
    V_\bk \equiv    \left(\begin{array}{cccc}
            u_{1\,\bk} & v_{1\,\bk} & \mu_{1\,\bk} & \nu_{1\,\bk} \\
            v_{1\,\bk} & u_{1\,\bk} & \nu_{1\,\bk} & \mu_{1\,\bk} \\
            u_{2\,\bk} & v_{2\,\bk} & \mu_{2\,\bk} & \nu_{2\,\bk} \\
            v_{2\,\bk} & u_{2\,\bk} & \nu_{2\,\bk} & \mu_{2\,\bk}
            \end{array}\right) \; .
\end{eqnarray}
The commutation relations~(\ref{Eq:ONRel}) are then written as 
\begin{eqnarray}\label{Eq:ONMat}
    V^{}_\bk g V^\tr_\bk = g \; , 
\end{eqnarray}
which implies $V^{}_\bk g V^\tr_\bk g = 1$, or
\begin{eqnarray}\label{Eq:InvV}
    V^{-1}_\bk & = & g V^\tr_\bk g
    \nonumber \\
               & = & \left(\begin{array}{rrrr}
            u_{1\,\bk} &   -v_{1\,\bk} &    u_{2\,\bk} &   -v_{2\,\bk} \\
           -v_{1\,\bk} &    u_{1\,\bk} &   -v_{2\,\bk} &    u_{2\,\bk} \\
          \mu_{1\,\bk} & -\nu_{1\,\bk} &  \mu_{2\,\bk} & -\nu_{2\,\bk} \\
         -\nu_{1\,\bk} &  \mu_{1\,\bk} & -\nu_{2\,\bk} &  \mu_{2\,\bk}
                     \end{array}\right) \; .   
\end{eqnarray}
This inverse transformation now allows us to state a convenient expression 
for the depletions: Since the many-particle ground state $|\text{gs}\rangle$ 
of the two-species gas is the quasiparticle vacuum,
\begin{eqnarray}
    \bo_{i\,\bk}|\text{gs}\rangle = 0  \quad \mbox{for} \; \bk \ne \bz \; ,
\end{eqnarray}
the occupation numbers $N_{i\,\bk}$ of the single-particle state with
wavevector~$\bk$ for the real particles of species~$i$, when the system
is in its ground state, are given by
\begin{subequations}\label{Eq:SingPartOcc}
\begin{eqnarray}
    \langle N_{1\,\bk}\rangle
    & = & \langle\text{gs}|\aoa_{1\,\bk}\ao_{1\,\bk}|\text{gs}\rangle
    \nonumber\\
    & = & \langle\text{gs}|(u_{1\,\bk}\boa_{1\,\bk}-v_{1\,\bk}\bo_{1\,-\bk}
          +u_{2\,\bk}\boa_{2\,\bk}-v_{2\,\bk}\bo_{2\,-\bk})
    \nonumber\\
    & \times & (u_{1\,\bk}\bo_{1\,\bk}-v_{1\,\bk}\boa_{1\,-\bk}
               +u_{2\,\bk}\bo_{2\,\bk}-v_{2\,\bk}\boa_{2\,-\bk})|\text{gs}\rangle
    \nonumber\\       
    & = & v_{1\,\bk}^2 + v_{2\,\bk}^2 \; ,
    \\
    \langle N_{2\,\bk}\rangle & = & \nu_{1\,\bk}^2 + \nu_{2\,\bk}^2  \; .
\end{eqnarray}
\end{subequations}
Therefore, the depletions are obtained from the sums 
\begin{subequations}\label{Eq:depletions}
\begin{eqnarray}
    N_1-N_{1\,\bz} &=& \sum_{\bk\ne\bz}\left(v_{1\,\bk}^2
                        +v_{2\,\bk}^2\right)   \;,\\
    N_2-N_{2\,\bz} &=& \sum_{\bk\ne\bz}\left(\nu_{1\,\bk}^2
                        +\nu_{2\,\bk}^2\right) \; ,
\end{eqnarray}
\end{subequations}
which will be evaluated in the following section. Contrary to the energy
shifts~(\ref{Eq:alpha}), the depletions do depend explicitly on the transformation
coefficients. 

The solution of the eigenvalue problem~(\ref{Eq:EigenProblem}) is elementary:  
Straightforward algebra leads to the eigenvalues   
\begin{eqnarray} \label{Eq:Spectra}
    \varepsilon_\pm^2(\bk) & = & 
    \frac{1}{2}\bigg[\varepsilon_{\rs1}^2(\bk) + \varepsilon_{\rs2}^2(\bk)
    \\ \nonumber & &
    \pm\sqrt{\left(\varepsilon_{\rs1}^2(\bk) - \varepsilon_{\rs2}^2(\bk)\right)^2
    + 16 t_1(\bk) t_2(\bk) v_{12}^2(\bk)} \bigg] \; ,
\end{eqnarray}
where here and in the following we write $\varepsilon_+(\bk)$ for the larger
of the two eigenvalues $\varepsilon_1(\bk)$ and $\varepsilon_2(\bk)$, and
$\varepsilon_-(\bk)$ for the smaller; moreover, $\varepsilon_{\rs i}(\bk)$
denotes the familiar quasiparticle energies of a single-species Bose gas,
\begin{eqnarray} \label{Eq:SingSpeSpec}
    \varepsilon_{\rs i}^2(\bk) 
    = t_{i}(\bk) \left[t_{i}(\bk) + 2v_{ii}(\bk)\right] \; .
\end{eqnarray}
As anticipated, with each eigenvalue of the problem~(\ref{Eq:EigenProblem}) 
also its negative appears; however, we only have to consider the positive ones. 
This spectrum~(\ref{Eq:Spectra}) has already been obtained through different 
reasoning in Refs.~\cite{GoldsteinMeystre97,AlexandrovKabanov02}. In the limit 
of vanishing interaction between the two species, $v_{12}(\bk)=0$, one sees that  
$\varepsilon_+(\bk)$ and $\varepsilon_-(\bk)$ become equal to the larger and the 
smaller of the single-species eigenvalues $\varepsilon_{\rs1}(\bk)$ and 
$\varepsilon_{\rs2}(\bk)$, respectively. Thus, without interspecies interaction 
the single-species solutions are recovered. ``Switching on'' $v_{12}(\bk)$ makes 
the upper and lower spectral branch repel each other: If the single-species 
spectra cross for some finite $\bk_0 \ne \bz$, so that 
$\varepsilon_{\rs1}(\bk_0) = \varepsilon_{\rs2}(\bk_0) \equiv \varepsilon_0$, 
one finds $\varepsilon_{\pm}^2(\bk_0) = 
\varepsilon_0^2\pm2[t_1(\bk_0)t_2(\bk_0)v_{12}^2(\bk_0)]^{1/2}$, 
so that the crossing of $\varepsilon_+(\bk)$ and $\varepsilon_-(\bk)$ is avoided 
as a consequence of the interspecies interaction.

For $\bk \to \bz$, both spectral branches become phononlike, {\em i.e.\/} 
proportional to $\hbar|\bk|$~\cite{GoldsteinMeystre97,AlexandrovKabanov02}:
\begin{equation}
	\varepsilon_\pm^2(\bk) \to c_\pm^2\hbar^2\bk^2 \; , 
\end{equation}
with sound velocities $c_+$ and $c_-$ given by
\begin{equation}
	c_\pm^2 \equiv \frac{1}{2}\left[c_{11}^2+c_{22}^2
		       \pm\sqrt{(c_{11}^2 - c_{22}^2)^2 + 4c_{12}^4}\right] \; .
\end{equation}
Here,
\begin{equation}
 	c_{ii}^2 = \frac{v_{ii}(\bz)}{m_i} = \frac{u_{ii}(\bz) \, n_i}{m_i}		  
\end{equation}
denote the squares of the well-known sound velocities for the 
single-species gases; in addition, we have introduced
\begin{equation}  		  
	c_{12}^4 = \frac{v_{12}^2(\bz)}{m_1 m_2} 
	         = \frac{u_{12}^2(\bz) \, n_1 n_2}{m_1 m_2} \; .
\end{equation} 

It is remarkable that the sign of $U_{12}(\br)$, and thus the answer to the
question whether the interspecies interaction is attractive or repulsive, 
does not enter into the spectrum~(\ref{Eq:Spectra}). In contrast, the 
intraspecies interaction has to be repulsive in order to guarantee the 
existence of a homogeneous single-species condensate.
 
While the existence of a honogeneous binary condensate requires that
both $\varepsilon_+(\bk)$ and $\varepsilon_-(\bk)$ be real, the smaller
eigenvalue $\varepsilon_-(\bk)$ actually is real only as long as the 
interspecies interaction is sufficiently weak, that is, under the
condition that both  
\begin{equation}\label{Eq:SingSpecStabRel}
   \varepsilon_{\rs 1}^2(\bk) > 0 \; , \quad
   \varepsilon_{\rs 2}^2(\bk) > 0 
\end{equation}
and
\begin{equation}\label{Eq:SpecStabRel}
    4v_{12}^2(\bk) < \left[t_1(\bk) + 2v_{11}(\bk) \right] 
                     \left[t_2(\bk) + 2v_{22}(\bk) \right] 
\end{equation}
be satisfied. 
Otherwise, our stationary approach does not have a physically meaningful
solution. The formal appearance of complex eigenvalues is related to an
instability leading to phase segregation: When $\varepsilon_-(\bk)$ touches 
the $|\bk|$-axis at some $\bk_0$, the corresponding mode can grow without cost of 
energy; the slightest excitation then makes the system collapse into inhomogeneous 
structures. The condition~(\ref{Eq:SingSpecStabRel}) implies that it is not 
possible to stabilize an unstable single-species condensate by adding particles 
of another species. Since the kinetic energy contributions $t_i(\bk)$ grow like 
$\bk^2$, while the potential contributions vanish for large~$|\bk|$, the 
stability condition~(\ref{Eq:SpecStabRel}) is most likely to be violated 
for small momenta, or large wavelength-excitations. Invoking the lowest-order
Born approximations 
\begin{equation}\label{Eq:a0}
	a^{(0)}_{ij} = \frac{m_{ij}}{2\pi\hbar^2}u_{ij}(\bz)   
\end{equation}
to the $s$-wave scattering lengths $a_{ij}$ for scattering events between 
particles of species~$i$ and~$j$, where
\begin{equation}
   m_{ij} \equiv m_i m_j / (m_i+m_j)
\end{equation}   
denotes the respective reduced mass, this stability 
condition~(\ref{Eq:SpecStabRel}) reduces for $\bk \to \bz$ to
\begin{equation}\label{Eq:StabRel}
    \left(\frac{a^{(0)}_{12}}{m_{12}}\right)^2
    < \frac{a^{(0)}_{11} a^{(0)}_{22}}{m_{11} m_{22}} \; .
\end{equation}
This inequality parallels the one known from mean-field studies of binary
condensates~\cite{AoChui98,AoChui00}. 

The transformation coefficients determining the quasiparticle operators
$\bo_{+\,\bk}$ and $\bo_{-\,\bk}$ follow from
\begin{eqnarray}\label{Eq:uvmunu}
    u_\pm^2 &=& (\varepsilon_\pm+t_1)^2
        \left(\varepsilon_\pm^2-\varepsilon_{\rs2}^2\right)t_2 \;x^2_\pm \; ,
	\nonumber \\
    v_\pm^2 &=& (\varepsilon_\pm-t_1)^2
        \left(\varepsilon_\pm^2-\varepsilon_{\rs2}^2\right)t_2 \;x^2_\pm \; ,
	\nonumber \\
    \mu_\pm^2 &=& (\varepsilon_\pm+t_2)^2
        \left(\varepsilon_\pm^2-\varepsilon_{\rs1}^2\right)t_1 \;x^2_\pm \; ,
	\nonumber \\
    \nu_\pm^2 &=& (\varepsilon_\pm-t_2)^2
        \left(\varepsilon_\pm^2-\varepsilon_{\rs1}^2\right)t_1 \;x^2_\pm \; ,
\end{eqnarray}
where we have omitted the index $\bk$ for clarity, and the normalization 
factors $x_+$ and $x_-$ are given by
\begin{eqnarray}
    \frac{1}{x_\pm^2} = 4t_1t_2 \varepsilon_\pm \left( 
    2\varepsilon_\pm^2 - \varepsilon_{\rs1}^2 - \varepsilon_{\rs2}^2 
    \right) \; ,
\end{eqnarray}
so that the proper commutation relations $[\bo_\pm\,,\boa_\pm] = 1$ are satisfied. 
The explicit knowledge of the coefficients $v_{\pm\,\bk}$ and $\nu_{\pm\,\bk}$ will 
be required in the following section for evaluating the depletions according to
Eqs.~(\ref{Eq:depletions}).

\section{Ground-state energy and depletion \label{Sec:GrStEn}}

From the definition~(\ref{Eq:ABCD1}) and the Born approximations~(\ref{Eq:a0}) 
we infer that the first contribution to the ground-state energy~(\ref{Eq:E0Formal}) 
can be written as
\begin{eqnarray}\label{Eq:a0Terms}
    A_\bz & = & 2\pi\hbar^2\left[ \frac{a^{(0)}_{11}n_1}{2m_{11}}N_1
                                + \frac{a^{(0)}_{22}n_2}{2m_{22}}N_2
    \right.\\ \nonumber & & \qquad\qquad\qquad\qquad\left.
    + \frac{a^{(0)}_{12}\sqrt{n_1n_2}}{m_{12}}\sqrt{N_1N_2} \right] \; .
\end{eqnarray}
As already emphasized, when performing the transformation from the 
Hamiltonian~(\ref{Eq:H_ABCD}) to the quasiparticle form~(\ref{Eq:HDem}), 
each wavevector~$\bk$ furnishes the additional contribution~(\ref{Eq:alpha}),
namely,
\begin{equation}\label{Eq:alphaPM}
    \alpha(\bk) = \frac{1}{2}\left[\varepsilon_+(\bk) +\varepsilon_-(\bk) 
		-\sum_{i=1}^2 \left[t_i(\bk)+v_{ii}(\bk)\right]\right] \; . 
\end{equation}
The evaluation of the sum over these shifts $\alpha(\bk)$ is notoriously difficult 
even in the case of a single-species condensate. If one replaces the $\bk$-dependent 
interaction matrix elements by a constant, the sum becomes ultraviolet divergent and 
requires careful regularization~\cite{PethickSmith02}; in the context of the 
pseudopotential method, this regularization is achieved by the derivative operator 
appearing in the proper pseudopotential~(\ref{Eq:PseudoR}). In standard textbook 
treatments some delicate technical details of this procedure are tacitly swept under 
the carpet through the identification of the full scattering length with only its 
first Born approximation~\cite{Landau-Stat2,Pathria-Stat}. However, the artificial 
divergence, and the subsequent need for regularization, can be avoided altogether
if the actual $\bk$-dependence of the interaction matrix elements is kept. Here we 
follow an approach recently developed in Ref.~\cite{WeissEtAl04}, and employ a 
density expansion for computing the sum in the low-density limit: Starting from 
Eq.~(\ref{Eq:alphaPM}) with the abbreviations~(\ref{Eq:tivij}), somewhat tedious 
but straightforward calculations yield 
\begin{eqnarray}
    \label{Eq:Expand1}
    \sum_{\bk\ne\bz}\alpha(\bk) \bigg|_{n_1,n_2=0}
    &=& 0 \; , \\
    \sum_{\bk\ne\bz}\frac{\partial\alpha(\bk)}{\partial n_i}\bigg|_{n_1,n_2=0}
    &=& 0 \; , \\
    \sum_{\bk\ne\bz}\frac{\partial^2\alpha(\bk)}{\partial n_i\partial n_j}
    	\bigg|_{n_1,n_2=0}
    &=& -\frac{1}{2}\sum_{\bk\ne\bz} \frac{u_{ij}^2(\bk)}{\hbar^2\bk^2/4m_{ij}} \; .
    \label{Eq:Expand3}
\end{eqnarray}
Now we observe that the next-to-leading term $a^{(1)}_{ij}$ of the Born series 
\begin{equation}
   a_{ij} = a^{(0)}_{ij} + a^{(1)}_{ij} + \ldots
\end{equation}   
for the various $s$-wave scattering lengths is given, in the limit of large 
volume~$V$, by~\cite{WeissEtAl04,CohenTannoudjiEtAl77}
\begin{equation}
    a^{(1)}_{ij} = 	
    -\left(\frac{m_{ij}}{2\pi\hbar^2}\right)^2 \frac{4\pi}{V}
    \sum_{\bk\ne\bz} \frac{u_{ij}^2(\bk)}{\bk^2} \; ,
\end{equation}
which implies
\begin{equation}
    \sum_{\bk\ne\bz} \frac{\partial^2\alpha(\bk)}{\partial n_i\partial n_j}
    \bigg|_{n_1,n_2=0} = \frac{2\pi\hbar^2}{m_{ij}} a_{ij}^{(1)} V \; . 
\end{equation}
Therefore, to second order in the densities the sum over the energy
shifts is given by
\begin{eqnarray}\label{Eq:a1Terms}
    \sum_{\bk\ne\bz} \alpha(\bk)
    & \simeq & 2\pi\hbar^2\left[
    \frac{a^{(1)}_{11}n_1}{2m_{11}}N_1+\frac{a^{(1)}_{22}n_2}{2m_{22}}N_2
    \right.\\ \nonumber && \qquad\qquad\qquad\left.
    + \frac{a^{(1)}_{12}\sqrt{n_1n_2}}{m_{12}}\sqrt{N_1N_2}\right] \; .
\end{eqnarray}
Adding Eqs.~(\ref{Eq:a0Terms}) and~(\ref{Eq:a1Terms}), we thus find the approximate
energy density for the ground state of the binary condensate,
\begin{eqnarray}\label{Eq:GrStEnLeading}
    \frac{E_0}{V} \simeq 2\pi\hbar^2\left[	
    \frac{\left(a^{(0)}_{11}+a^{(1)}_{11}\right)}{2m_{11}} n_1^2
  + \frac{\left(a^{(0)}_{22}+a^{(1)}_{22}\right)}{2m_{22}} n_2^2 \right.
    \nonumber \\ \qquad\qquad \left. 
  + \frac{\left(a^{(0)}_{12}+a^{(1)}_{12}\right)}{m_{12}} n_1 n_2 \right] \; .
\end{eqnarray}
At this point, a major weakness of the quadratic approximation employed for 
the reduction of the orignal Hamiltonian~(\ref{Eq:OrigH}) becomes apparent: 
Although the contribution~(\ref{Eq:a0Terms}) arises from the interaction term 
with all four wavevector-indices equal to the condensate index $\bz$, and the 
contribution~(\ref{Eq:a1Terms}) stems from terms supposed to be ``quadratically 
small'', they are of the {\em same\/} order in the densities. The intuitive 
categorization of the interaction terms in the original Hamiltonian~(\ref{Eq:OrigH}) 
with respect to the number of ground-state operators they contain thus does not
reflect a systematic ordering. The ``quadratic'' contributions to the Hamiltonian 
are small compared to the leading one, $A_\bz$, only if the Born series for the 
scattering lengths converge rapidly, so that 
\begin{equation}
   \label{Eq:ineq}
   |a_{ij}^{(1)}| \ll |a_{ij}^{(0)}| \; ,
\end{equation}
implying that the first terms $a_{ij}^{(0)}$ already provide good approximations 
to the full scattering lengths $a_{ij}$. 
 
If one accepts the approximation~(\ref{Eq:GrStEnLeading}), one can derive
a criterion for phase segregation by comparing the ground-state 
energy~(\ref{Eq:GrStEnLeading}) of a homogeneous mixture of $N_1$ and $N_2$
Bose particles in a volume~$V$ with that of two single-species condensates 
occupying separate volumes $V_1$ and $V_2$ with $N_1$ and $N_2$ particles,
respectively, while maintaining a constant total volume $V = V_1 + V_2$. 
Neglecting surface effects, the mixture is stable as long as its ground-state 
energy is less than the lowest possible value the sum of two separate ground-state 
energies can attain for any partition $V_1 + V_2 = V$~\cite{AoChui98}. Performing 
the minimization, this stability requirement is translated into the condition
\begin{equation}\label{Eq:MFstab}
    \left( \frac{a^{(0)}_{12}+a^{(1)}_{12}}{m_{12}}\right)^2
    < \frac{\left(a^{(0)}_{11}+a^{(1)}_{11}\right)
            \left(a^{(0)}_{22}+a^{(1)}_{22}\right)}{m_{11}m_{22}} \; . 
\end{equation}
Since we require the inequalities~(\ref{Eq:ineq}) for consistency, this stability
condition is practically the same as the condition~(\ref{Eq:StabRel}) deduced from
the requirement of real quasiparticle energies.

Continuing with the density expansion of the ground-state energy, the next-order
contribution turns out to be of intriguing complexity. While our general scheme
is fully capable of treating the general case, and one may use this scheme to generate 
the corresponding terms with the help of symbolic algebraic manipulation software, 
we restrict ourselves here to the experimentally relevant case of equal masses,
$m_1 = m_2 \equiv m$ or $t_1 = t_2 \equiv t$, as corresponding, {\em e.g.\/}, to a 
binary condensate consisting of atoms of a single isotope in two different spin 
states~\cite{MyattEtAl97,StamperKurnEtAl98,HallEtAl98,MatthewsEtAl99,MaddaloniEtAl00}.
The resulting expressions then simplify considerably: The quasiparticle energies 
$\varepsilon_\pm(\bk)$ acquire a form which is strongly reminiscent of the 
single-species expressions~(\ref{Eq:SingSpeSpec}), 
\begin{equation}\label{Eq:EqualMassSpec}
	\varepsilon_\pm^2(\bk) = t(\bk)[t(\bk)+2u_\pm(\bk) n] \; .
\end{equation}
Here we have introduced the auxiliary functions  
\begin{eqnarray}
	u_\pm(\bk) & \equiv & 
	\frac{1}{2}\bigg[c_1u_{11}(\bk)+c_2u_{22}(\bk)
	\\\nonumber & & 
	\pm \sqrt{[c_1u_{11}(\bk)-c_2u_{22}(\bk)]^2+4c_1c_2u_{12}^2(\bk)}\bigg] \; , 
\end{eqnarray}
which depend not only on the three interaction potentials, but also on the
concentrations $c_1$ and $c_2$ of the individual species. We also write 
$n = n_1 + n_2$ for the total density, implying both $n_i = c_i n$ and
$c_1 + c_2 = 1$. For vanishing interspecies interaction, $u_{12}(\bk)=0$, 
one finds that $u_+(\bk)$ and $u_-(\bk)$ become equal to the maximum and the
minimum of $c_1u_{11}(\bk)$ and $c_2u_{22}(\bk)$, respectively. The explicit 
appearance of the total density~$n$ in Eq.~(\ref{Eq:EqualMassSpec}) greatly 
faciliates the following calculations.  

It seems to be tempting now to go on with the Taylor expansion~(\ref{Eq:Expand1})
-- (\ref{Eq:Expand3}) in powers of the density, and to compute
\begin{eqnarray}
	\sum_{\bk\ne\bz} \frac{\partial^3\alpha(\bk)}{\partial n^3}\bigg|_{n=0}
	= \frac{3}{2}\sum_{\bk\ne\bz}\frac{u_+^3(\bk)+u_-^3(\bk)}{t^2(\bk)} \; ,
\end{eqnarray}
but this expression obviously becomes infrared divergent in the thermodynamic
limit, and therefore is of no use here. This finding reflects the fact that, 
analogous to the single-species case, the proper expansion parameters for
the ground-state energy of a dilute binary Bose gas are not the parameters
$n_i a_{ij}^3$ themselves, but their square roots. Correspondingly, the expression
\begin{eqnarray}
	& & \sum_{\bk\ne\bz} 
	\frac{\partial}{\partial\sqrt{n}}\frac{\partial^2\alpha(\bk)}{\partial n^2}
	\bigg|_{n=0} =  
	\\ \nonumber & & 
	\lim_{n\to 0} \frac{V}{(2\pi)^3}\int \! \rd\bk  
	 \left[\frac{3(tu_+)^3\sqrt{n}}{\left(t^2+2tu_+n\right)^{5/2}}
	      +\frac{3(tu_-)^3\sqrt{n}}{\left(t^2+2tu_-n\right)^{5/2}} \right]  
\end{eqnarray}
remains finite~\cite{WeissEtAl04}. For evaluating these integrals, we introduce 
a dimensionless momentum variable~$x$ by demanding
\begin{equation} 
   \label{Eq:x} 
   t = \hbar^2 \bk^2/2m \equiv  u n x^2 \; ,
\end{equation}
with a constant~$u$ carrying the same dimension as $u_\pm(\bk)$, energy times
volume. The limit $n \to 0$ can then be taken under the integral, implying
$u_\pm(\bk) \to u_\pm(\bz)$ there, and thus leading to
\begin{eqnarray}
	& & \sum_{\bk\ne\bz} 
	\frac{\partial}{\partial\sqrt{n}}\frac{\partial^2\alpha(\bk)}{\partial n^2}
	\bigg|_{n=0} = 
	\\ \nonumber & & 
	\frac{4\pi V}{(2\pi\hbar)^3} (2mu)^{3/2} \, 3u  
	\int_{0}^\infty \! \rd x \sum_{+,-}
	\frac{(u_\pm(\bz)/u)^3 x^3}{\left(x^2 + 2u_\pm(\bz)/u\right)^{5/2}} \; .
\end{eqnarray}
With the help of the identity 
\begin{equation}
   \int_0^\infty \! \rd x \, \frac{x^3}{\left(x^2 + c\right)^{5/2}}
   = \frac{2}{3\sqrt{c}}
\end{equation}
one then finds 
\begin{equation}
	\sum_{\bk\ne\bz} 
	\frac{\partial}{\partial\sqrt{n}}\frac{\partial^2\alpha(\bk)}{\partial n^2}
 	\bigg|_{n=0} =  
	\frac{2m^{3/2}V}{\pi^2\hbar^3} \sum_{+,-} u_\pm(\bz)^{5/2} \; .
\end{equation}
Introducing the scattering length-like quantities
\begin{eqnarray}
	a^{(0)}_\pm & \equiv & \frac{m}{4\pi\hbar^2} u_\pm(\bz)
	\\ \nonumber
	& = &\frac{1}{2}\bigg[c_1a^{(0)}_{11}+c_2a^{(0)}_{22}
	\\\nonumber && \qquad
	\pm \sqrt{ \left[c_1a^{(0)}_{11}-c_2a^{(0)}_{22}\right]^2
	    + 4c_1c_2{a^{(0)}_{12}}^2} \, \bigg] \; ,
\end{eqnarray}	
which, however, again depend explicitly on the concentrations $c_1$ and $c_2$ of the
individual species, the next contribution to the ground-state energy acquires the form
\begin{eqnarray}\label{Eq:Beyond}	
	& & \left.\sum_{\bk\ne\bz} \frac{\partial}{\partial\sqrt{n}}
	\frac{\partial^2\alpha(\bk)}{\partial n^2}\right|_{n=0}
	\frac{2}{5}\frac{2}{3}n^{5/2}
	\nonumber  \\ & = & 
	\frac{2\pi\hbar^2}{m} n^2 V \frac{128}{15\sqrt{\pi}}
	\sum_{+,-} a^{(0)}_\pm \sqrt{{n a^{(0)}_\pm}^3} \; .
\end{eqnarray}
Hence, adding this result to the previous contribution~(\ref{Eq:GrStEnLeading}), 
we obtain for the ground-state energy density of a homogeneous binary condensate, 
consisting of two types of Bosons with equal masses, the expression 
\begin{eqnarray}
       \label{Eq:Eden}
       \frac{E_0}{V} &=& \frac{2\pi\hbar^2}{m} n^2 \Bigg[	
       c_1^2\left(a^{(0)}_{11}+a^{(1)}_{11}\right)
       +c_2^2\left(a^{(0)}_{22}+a^{(1)}_{22}\right)
       \nonumber \\ & & 
       + \,2c_1c_2\left(a^{(0)}_{12}+a^{(1)}_{12}\right)
       \nonumber \\ & & 
       + \frac{128}{15\sqrt{\pi}}\sum_{+,-} a^{(0)}_\pm 
       \sqrt{{n a^{(0)}_\pm}^3} \Bigg] \; .
\end{eqnarray} 
In the limiting case of vanishing interspecies interaction one has
\begin{eqnarray}
	\label{Eq:Limit}
	a^{(0)}_+ & \to & \max\{ c_1a^{(0)}_{11}, c_2 a^{(0)}_{22} \} \; ,
	\nonumber \\
	a^{(0)}_- & \to & \min\{ c_1a^{(0)}_{11}, c_2 a^{(0)}_{22} \} \; ,	
\end{eqnarray}
giving
\begin{equation}
        \label{Eq:Elim}
	\frac{E_0}{V} = \sum_{i=1}^2 
	\frac{2\pi\hbar^2}{m} n_i^2\left[a^{(0)}_{ii}+a^{(1)}_{ii}
	+ a^{(0)}_{ii}\frac{128}{15\sqrt{\pi}} \sqrt{{n_i a^{(0)}_{ii}}^3} \right] \; .
\end{equation} 
Since we are relying on the assumption that the Born series for the scattering 
lengths converge rapidly, so that $a \simeq a^{(0)}_{ii} + a^{(1)}_{ii} 
\simeq a^{(0)}_{ii}$, this formula~(\ref{Eq:Elim}) reduces properly to the expected 
sum of two Lee-Yang-type expressions~(\ref{Eq:LY}).

Finally, the calculation of the depletions according to Eqs.~(\ref{Eq:depletions})
proceeds along similar lines. For species~1 we have to evaluate
\begin{equation}
	N_{1} - N_{1\,\bz} 
	= \sum_{\bk\ne\bz}\left(v_{+\,\bk}^2+v_{-\,\bk}^2\right) \; ,
\end{equation}
with the transformation coefficients $v_{\pm\,\bk}$ read off from Eqs.~(\ref{Eq:uvmunu}). 
Restricting ourselves again to the case of equal masses, and employing once more the 
continuous dimensionless variable~$x$ introduced in Eq.~(\ref{Eq:x}) for converting the 
sum over the discrete wavevectors~$\bk$ in the thermodynamic limit into an integral, 
we find
\begin{eqnarray}
	v_{\pm\,\bk}^2 & = & 
	\frac{(\varepsilon_\pm - t)^2
	\left(\varepsilon_\pm^2 - \varepsilon_{\rs2}^2\right)}
        {4\varepsilon_\pm t \left(2\varepsilon_\pm^2 - \varepsilon_{\rs1}^2
                                                     - \varepsilon_{\rs2}^2 \right)} 
	\\  \nonumber & = &
	\frac{1}{2} \frac{u_\pm - c_2 u_{22}}{2u_\pm - c_1 u_{11} - c_2 u_{22}}
	\left[ \frac{x^2 + u_{\pm}/u}{x\sqrt{x^2 + 2 u_{\pm}/u}} - 1 \right] \; .		
\end{eqnarray}
Taking the limit $n \to 0$ in the same manner as above under the integral, we obtain 
after some manipulations the desired depletion    
\begin{subequations}\label{Eq:dep}
\begin{eqnarray}
	\frac{N_{1} - N_{1\,\bz}}{V}  & \simeq & \frac{8}{3\sqrt{\pi}} n
	\\ \nonumber & \times & \sum_{+,-}
	\frac{a^{(0)}_\pm- c_2 a^{(0)}_{22}}
	     {2a^{(0)}_\pm-c_1a^{(0)}_{11}-c_2a^{(0)}_{22}} \,
	\sqrt{{n a^{(0)}_\pm}^3} \; ;
\end{eqnarray}
analogously, 
\begin{eqnarray}
	\frac{N_{2} - N_{2\,\bz}}{V} & \simeq & \frac{8}{3\sqrt{\pi}} n
	\\ \nonumber & \times & \sum_{+,-}
	\frac{a^{(0)}_\pm-c_1a^{(0)}_{11}}
	     {2a^{(0)}_\pm-c_1a^{(0)}_{11}-c_2a^{(0)}_{22}} \,
	\sqrt{{n a^{(0)}_\pm}^3} \; .
\end{eqnarray}
\end{subequations}
The well-known single-species expression is easily recovered in the limit of 
vanishing interspecies interaction, as it should: Utilizing Eqs.~(\ref{Eq:Limit}), 
the above sums reduce to a single term, and we are left with
\begin{equation}
	\frac{N_i-N_{i\,\bz}}{V} \simeq 
	\frac{8}{3\sqrt{\pi}} n_i \sqrt{{n_i a^{(0)}_{ii}}^3 } \; .
\end{equation}

\section{Conclusion \label{Sec:Conclusion}}

The preceding calculations explicitly refer to soft interaction potentials
$U_{ij}(\br)$ with a well-behaved Fourier transform, possessing a rapidly
converging Born series for the respective scattering length, as is the case for wide 
and shallow potentials. In that case the first Born approximations $a_{ij}^{(0)}$ 
practically exhaust the full Born series. Since the functional dependence of at least 
the leading terms of the density expansion of the ground-state energy on the scattering 
lengths is the same for both soft interactions and hard ones with a short-range 
singularity~\cite{Landau-Stat2} (see also Ref.~\cite{ChernyShanenko00} for a recent 
discussion of this point), we may free ourselves from this restriction by replacing 
$a_{ij}^{(0)}$, or $a_{ij}^{(0)} + a_{ij}^{(1)}$, by the full scattering lengths: 
Given a binary condensate of two species with equal masses~$m$, total density~$n$, 
and species concentrations $c_1$ and $c_2$, interacting via some potentials 
$U_{ij}(\br)$ with scattering lengths $a_{ij}$, 
we define
\begin{eqnarray}
	a_\pm & \equiv & \frac{1}{2}\bigg[c_1 a_{11}+c_2 a_{22}
	\\\nonumber && \qquad
	\pm \sqrt{ \left[c_1 a_{11} - c_2a_{22} \right]^2
	         + 4 c_1 c_2 a_{12}^2} \, \bigg] \; .
\end{eqnarray}	
If the mixture is sufficiently dilute, such that 
\begin{equation}
 	\sqrt{n a_\pm^3} \ll 1 \; ,
\end{equation}
the ground-state energy density is then given by
\begin{eqnarray}
       \frac{E_0}{V} & = & \frac{2\pi\hbar^2}{m} n^2 \Bigg[	
       c_1^2 \, a_{11} + c_2^2 \, a_{22} + 2 c_1 c_2 \, a_{12}
       \nonumber \\ & & 
       + \frac{128}{15\sqrt{\pi}} 
       \left( a_+\sqrt{n a_+^3} + a_- \sqrt{n a_-^3} \right) \Bigg] \; , 
\label{Eq:Res1}
\end{eqnarray} 
while the depletion can be written as
\begin{eqnarray}
	\frac{N_{1} - N_{1\,\bz}}{V}  & = & \frac{8}{3\sqrt{\pi}} n
	\Bigg[ \frac{a_+ - c_2 a_{22}}{2a_+ - c_1 a_{11} - c_2 a_{22}} 
	\sqrt{n a_+^3} 
	\nonumber \\ & & \quad 
	     + \frac{a_- - c_2 a_{22}}{2a_- - c_1 a_{11} - c_2 a_{22}}  
	\sqrt{n a_-^3} \Bigg]  			
\label{Eq:Res2}
\end{eqnarray}
for species~$1$; for the other species the depletion is obtained by interchanging 
the indices~$1$ and~$2$.

The particular case $a_{12}^2 = a_{11} a_{22}$ implies $a_- = 0$, so that, according 
to Eq.~(\ref{Eq:EqualMassSpec}), one also has $\varepsilon_-(\bk) = t(\bk)$, 
independent of the densities. Hence, the lower branch of the spectrum then is 
collisionless; this phenomenon is of particular interest in atom 
interferometry~\cite{GoldsteinEtAl00}. If one disregards the contributions to the 
ground-state energy obtained in Eq.~(\ref{Eq:Beyond}), an argument analogous to 
the one which led to the stability criterion~(\ref{Eq:MFstab}) requires 
$a_{12}^2 < a_{11} a_{22}$ for a homogeneous binary condensate of particles with 
equal masses, and thus places a system with $a_- = 0$ right on the borderline 
between the regime of the homogeneous mixture and that of two segregated phases. 
However, inclusion of these contributions~(\ref{Eq:Beyond}) into the analysis shifts 
a system with $a_- = 0$ slightly into the regime where the mixture is stable: The 
contributions to the ground-state energy beyond those given by the mean-field 
approach tend to stabilize a homogeneous binary condensate with a collisionless 
branch of the quasiparticle spectrum.

Even though derived under the restriction of equal masses of both species, both 
the result~(\ref{Eq:Res1}) for the ground-state energy and, in particular, the 
expression~(\ref{Eq:Res2}) for the depletion is surprisingly involved and could not 
have been guessed on the basis of the known single-species formulas. The corresponding
expressions for the general case, which are substantially more cumbersome, are 
obtained by following exactly the steps explained above. 

Our reasoning clearly indicates that the elementary Bogoliubov theory, which 
works directly with the interparticle potentials, gets worse the deeper the 
potentials~\cite{Lieb63}, as the missing terms of the Born series become increasingly 
more important. In the case of a single-species condensate, the emergence of the 
full scattering length~$a$ instead of its first Born approximation in the 
expression~(\ref{Eq:LY}) for the ground-state energy density is justified by means 
of a summation of ladder diagrams~\cite{FetterWalecka03}, or by employing effective
field theory with the exact two-point scattering amplitude~\cite{Andersen04}; these
devices may also be used in the binary case. It should be kept in mind, however,  
that so far only the leading term in Eq.~(\ref{Eq:LY}) could be established with 
full mathematical rigor~\cite{LiebYngvason98}, whereas a corresponding rigorous 
proof for the next-to-leading term is still lacking. It would, therefore, be of 
considerable conceptual value to demonstrate in explicit detail how the missing terms 
of the Born series are recovered if one systematically takes into account those
contributions to the full Hamiltonian which have been neglected in the quadratic 
Bogoliubov approximation, thus showing the consistency of the various approaches. 
We will present such a calculation in a subsequent paper~\cite{WeissEckardt04}.      

\bibliography{bosegas}
\end{document}